\documentclass[twocolumn,prb,amsmath,amssymb,showpacs]{revtex4}
\usepackage{graphicx}
\pdfoutput=1

\begin{document}

\title{Theory of Spin Magnetohydrodynamics}

\author{Yaroslav Tserkovnyak}
\author{Clement H. Wong}
\affiliation{Department of Physics and Astronomy, University of California, Los Angeles, California 90095, USA}

\date{\today}

\begin{abstract}
We develop a phenomenological hydrodynamic theory of coherent magnetic precession coupled to electric currents. Exchange interaction between electron spin and collective magnetic texture produces two reciprocal effects: spin-transfer torque on the magnetic order parameter and the Berry-phase gauge field experienced by the itinerant electrons. The dissipative processes are governed by three coefficients: the ohmic resistance, Gilbert damping of the magnetization, and the ``$\beta$ coefficient" describing viscous coupling between magnetic dynamics and electric current, which stems from spin mistracking of the magnetic order. We develop general magnetohydrodynamic equations and discuss the net dissipation produced by the coupled dynamics. The latter in particular allows us to determine a lower bound on the magnetic-texture resistivity.
\end{abstract}



\maketitle

\section{Introduction}

Conduction electrons moving in a ferromagnet interact with the magnetization through the exchange interaction.  If the exchange field is strong and slowly varying in space and time, the electron spin will adiabatically follow the direction of the magnetization.  We may then consider electrons with spins up and down along the magnetization direction as two distinct species of particles, and for convenience call them spin up/down electrons.  As is well known, a spin up/down electron wave packet acquires a Berry phase\cite{berryPRSLA84} that influences their orbital motion. In effect, the electrons experience a Lorentz force due to ``fictitious" electromagnetic fields which are local functions of the magnetization.\cite{volovikJPC87}

In this fictitious electrodynamics, spin up/down electrons have opposite charges and different conductivities. Their motion and associated currents interact with the magnetization through what is commonly called current-driven spin-transfer torques.  We call this interplay between spin currents and magnetization \textit{spin magnetohydrodynamics}, in analogy to the classical theory of magnetohydrodynamics,\cite{landauBOOKv8} where the magnetic fields couple to electric currents in conducting fluids, and the currents in turn generate magnetic fields.  In our spin magnetohydrodynamics, the Maxwell's equations for the magnetic field are replaced by the Landau-Lifshitz-Gilbert (LLG) equation for the magnetization.  In this paper, we neglect full dynamics of the real electromagnetic fields, focusing on the spin-related phenomena. 

The electron spin follows the magnetization direction perfectly only in the limit of an infinitely large exchange field.   In reality, there will be some misalignment and associated spin relaxation.  This is usually described phenomenologically as a dissipative spin torque with a coefficient $\beta$ in the Landau-Lifshitz equation.\cite{zhangPRL04,thiavilleEPL05,tserkovPRB06md} In a one-dimensional ring geometry, we will derive the complete set of coupled spin-magnetohydrodynamical equations, starting from the semi-phenomenological dynamical equations for nonequilibrium currents and magnetization.  We recast the reactive spin torque mediated by the Berry phase in this thermodynamic context.   In our theory, we take an alternative view that the $\beta$ term arises from a correction to the Berry-phase electromotive force (EMF) in the equation of motion for the charge current, with the appropriate dissipative spin torque established by the Onsager reciprocity.

This physics is presently vigorously studied (experimentally as well as theoretically) in the contexts of current-driven magnetic excitations and domain-wall motion\cite{yamaguchiPRL04,zhangPRL04,thiavilleEPL05,barnesPRL05,tserkovPRB06md,kohnoJPSJ07,yamanouchiSCI07} and the reciprocal spin accumulations and voltages generated by the fictitious gauge fields.\cite{barnesPRL07,saslowPRB07,duinePRB08,tserkovPRB08,moriyamaPRL08,tserkovCM08tb} Since the mesoscopic regime (mainly dealing with variants of magnetic spin valves, tunnel junctions, and magnetic multilayers) is at present well explored,\cite{tserkovRMP05} we will limit our attention here to the case of continuous magnetic systems.

\section{Nondissipative spin torque}

\begin{figure}
\centerline{\includegraphics[width=\linewidth]{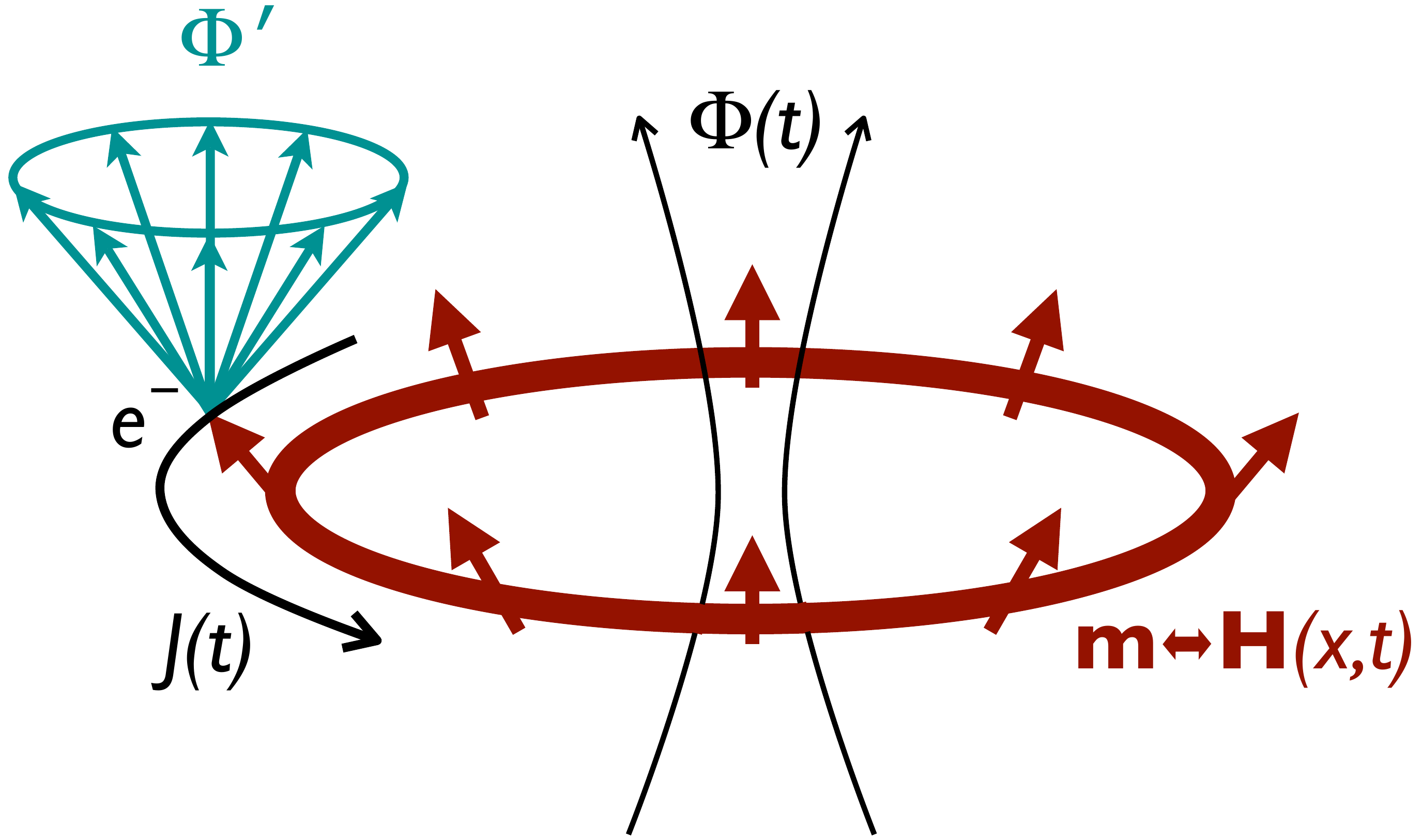}}
\caption{(color online). Schematics of our principal ``study case:" Uniform electric current $J(t)$ carried by itinerant electrons can be driven by the external magnetic flux $\Phi(t)$ generating the EMF $\mathcal{E}=-\partial_t\Phi/c$. The magnetic texture $\mathbf{m}(x,t)$ responds to the effective field $\mathbf{H}(x,t)$, which may have an external contribution applied to the wire independently of $\Phi$. The reactive magnetohydrodynamic coupling stems from the Berry phase $\Phi^\prime$, which is acquired by the electron spin (shown in blue) following the instantaneous magnetic profile (shown in red) around the loop. $\Phi^\prime$ corresponds geometrically to the solid angle enclosed by the electron spin. Coupled dissipative processes arise once we relax the projection approximation, allowing for some orientational spin mistracking and dephasing as electrons propagate through the magnetic texture.}
\label{fig}
\end{figure}

Since the underlying physics is rich and complex in the most general setting, we will limit our discussion to a simple setting, which we believe captures all the essential ingredients of the spin magnetohydrodynamics.  Consider a uniform current in a ferromagnetic ring, assuming for simplicity incompressible electric flows (the continuity equation prohibits current inhomogeneities for an incompressible electron fluid).  The electric current is then the only dynamical variable describing the electron fluid.  The magnetic texture here could be a domain wall or  magnetic spiral, for example (in higher dimensions we could have topological twists and kinks such as vortices, hedgehogs, or skyrmions).  See Fig.~\ref{fig} for a schematic of the setup.  In the Landau-Lifshitz phenomenology of ferromagnetic dynamics well below the Curie temperature, only the instantaneous direction of the magnetization $\mathbf{m}(x,t)$ (or, equivalently, spin density) is assumed to be a dynamic variable. The magnitude of the spin density $S$ along $\mathbf{m}$ is assumed to be uniform and constant in time.   We will separately drive the current with a time-dependent external magnetic flux $\Phi(t)$ inside the ring, and the magnetic dynamics with a magnetic field $\mathbf{h}(x,t)$ applied directly to the wire.

The first step in our phenomenology is to identify the free energy $\mathcal{F}$ as a function of the thermodynamic variables $J$ and $\mathbf{m}(x,t)$ (or their thermodynamic conjugates), which completely determine the macroscopic state of our system, assuming local thermal equilibrium. Neglecting spin, the gauge-invariant free energy associated with an electric current in the ring is given by $\mathcal{F}(\mathcal{J},\Phi)=\left(\mathcal{J}-\Phi/c\right)^2/2L$, where we define $L\mathcal{J}$ to be the current corresponding to the \textit{canonical} momentum of the electrons.  $L$ is the self-inductance of the ring and  $c$ is the speed of light. However, spin up/down electrons propagating through a quasistatic magnetic texture\cite{sternPRL92} accumulate also a Berry phase,\cite{berryPRSLA84} which gives a fictitious contribution to the vector potential associated with a fictitious EMF.\cite{barnesPRL07} This vector potential is given (in some convenient gauge) by\cite{tserkovPRB08} $A^\prime_x=(\hbar c/e)\sin^2(\theta/2)\partial_x\phi$,  producing gauge-invariant fictitious flux,
\begin{equation}
\Phi^\prime=\oint dx A^\prime_x =\frac{\hbar c}{2e}\oint dx(1-\cos\theta)\partial_x\phi\,.
\label{phi}
\end{equation}
$(\theta,\phi)$ are the spherical angles parametrizing $\mathbf{m}(x)$. $e>0$ is minus the electron charge. Eq.~(\ref{phi}) is the flux associated with spin-up electrons adiabatically following magnetic texture, with the opposite result for spin-down electrons.  

The free energy accounting for the Berry phase becomes
\begin{equation}
\mathcal{F}^\prime(\mathcal{J}, \Phi, \Phi^\prime [\mathbf{m}(x,t)] )=\left[\mathcal{J}-(\Phi+p\Phi^\prime)/c\right]^2/2L\,,
\end{equation}
where  $p$ is the polarization of the spin $s$-dependent conductivity $\sigma_s$: $p=(\sigma_\uparrow-\sigma_\downarrow)/(\sigma_\uparrow+\sigma_\downarrow)$ (assuming fast spin relaxation or halfmetallic ferromagnets). The electric current is given by
\begin{equation}
J\equiv-c\partial_\Phi\mathcal{F}^\prime=\left[\mathcal{J}-(\Phi+p\Phi^\prime)/c\right]/L=\partial_\mathcal{J}\mathcal{F}^\prime\,,
\end{equation}
which is thus the thermodynamic conjugate of $\mathcal{J}$. The equation of motion for current in our simple electric circuit is given by Ohm's law,
\begin{equation}
\partial_t\mathcal{J}\equiv L\partial_tJ+\partial_t(\Phi+p\Phi^\prime)/c=-RJ \,.
\label{ohm}
\end{equation}
 where $R$ is the resistance of the wire. Naturally, the dynamic Berry phase is seen to give a contribution to the EMF:\cite{barnesPRL07}
\begin{equation}
\mathcal{E}^\prime\equiv-p\partial_t\Phi^\prime/c=\mathcal{P}\oint dx\,\mathbf{m}\cdot(\partial_x\mathbf{m}\times\partial_t\mathbf{m})\,,
\label{EMF}
\end{equation}
which is a well-known result.\cite{volovikJPC87} (We defined $\mathcal{P}=p\hbar/2e$.)

Now that the free energy of the current is coupled to the magnetization of the ring through the Berry-phase flux, there will be a corresponding reactive coupling of the magnetization to the current.  We describe magnetic dynamics by the Landau-Lifshitz-Gilbert equation\cite{landauBOOKv9}
\begin{equation}
\partial_t\mathbf{m}=\mathbf{H}\times\mathbf{m}/S-\alpha\mathbf{m}\times\partial_t\mathbf{m}\,,
\label{LLG}
\end{equation}
where the effective field $\mathbf{H}$ is defined by the functional derivative, $\mathbf{H}\equiv\partial_{\mathbf{m}}\mathcal{F}$ (so that locally $\mathbf{H}\perp\mathbf{m}$), and $\alpha$ is the dimensionless Gilbert damping\cite{gilbertIEEEM04} parameter. The total free energy of our magnetoelectric system is $\mathcal{F}(\mathbf{m}, \mathcal{J},\Phi)=\mathcal{F}(\mathbf{m})+\mathcal{F}^\prime(\mathcal{J}, \Phi, \Phi^\prime[\mathbf{m}(x,t)] )$, where $\mathcal{F}(\mathbf{m})$ is a standard free energy of the ferromagnet. Variation of the $\mathcal{F}^\prime$  with respect to $\mathbf{m}$ gives current-driven spin torque applied to the magnetic dynamics:\cite{bazaliyPRB98} $\boldsymbol{\tau}^\prime\equiv\partial_\mathbf{m}\mathcal{F}^\prime\times\mathbf{m}$, where $\partial_\mathbf{m}\mathcal{F}^\prime\equiv\partial_{\mathbf{m}}\Phi^\prime\partial_{\Phi^\prime}\mathcal{F}^\prime=-pJ\partial_{\mathbf{m}}\Phi^\prime/c$. Differentiating Berry phase (\ref{phi}) with respect to $\mathbf{m}$, we find
\begin{equation}
\boldsymbol{\tau}^\prime= \mathcal{P}J\partial_x\mathbf{m}\,.
\label{tau}
\end{equation}
Since $\hbar/2e$ is the electron spin-charge conversion factor, we can give another interpretation of this term.  It is simply the rate of change of the angular momentum of the conducting electrons with spins locked to the magnetic profile.  The spins of the up/down electrons rotate in the opposite directions so that, if the spin up/down conductivities are the same (and thus $\mathcal{P}=0$), the net change in their angular momentum vanishes.  Putting this term on the left-hand side, we get
\begin{equation}
\partial_t\mathbf{m}-\mathcal{P}J\partial_x\mathbf{m}/S=\partial_\mathbf{m}\mathcal{F}(\mathbf{m})\times\mathbf{m}/S-\alpha\mathbf{m}\times\partial_t\mathbf{m}\,.
\end{equation}
The left-hand side of this equation is the rate of change of the total angular-momentum density of the magnetoelectric system,\cite{volovikJPC87} while the right-hand side gives the usual LLG torque on the system.

\section{Dissipative spin torque}

LLG equation (\ref{LLG}) with torque (\ref{tau}) and Ohm's law (\ref{ohm}) with the fictitious EMF (\ref{EMF}) now constitute coupled equations of our spin magnetohydrodynamic theory, with the reactive coupling mediated by Berry phase (\ref{phi}). We reproduce them here for clarity (after putting the magnetization equation in the Landau-Lifshitz form):
\begin{equation}
\partial_t\mathcal{J}=-RJ\,,\,\,\,\,\,\,\partial_t\mathbf{m}=\frac{\mathbf{H}\times\mathbf{m}-\alpha\mathbf{H}}{(1+\alpha^2)S}\,.
\label{LL}
\end{equation}
These are the equations of motion for a quasistationary, thermodynamic system near equilibrium.\cite{landauBOOKv5} In equilibrium, the current $J$ is zero and magnetization is static. Out of equilibrium, the first-order time derivatives of $(\mathcal{J},\mathbf{m})$ are completely specified by the instantaneous values of their thermodynamic conjugates $(J,\mathbf{H})$. The right-hand side is a linear expansion in these conjugates with dissipative coefficients $R$ and $\alpha$ that cause the system to relax back to equilibrium.  So far, the dissipation in the current and magnetization is separate and physically unrelated.  We now add the dissipative couplings which will be key results of this paper.    

We proceed phenomenologically by adding to the current equation (\ref{ohm}) correction  $\Delta\mathcal{E}^\prime$ to the Berry-phase EMF and correction $R^\prime$ to resistance, due to coupling with the magnetic texture $\mathbf{m}(x,t)$. The modified Ohm's law then becomes:
\begin{equation}
\partial_t\mathcal{J}=-(R+R^\prime)J+\Delta\mathcal{E}^\prime
\label{ohmD}
\end{equation}
To avoid a slew of uninteresting coefficients and anisotropies, we will constrain the phenomenology by assuming spin-rotational symmetry of the magnetic texture and the inversion symmetry of the wire. Under the latter, $\mathbf{m}\to\mathbf{m}$, $J\to-J$, $\partial_x\to-\partial_x$, and $\mathcal{E}^\prime\to-\mathcal{E}^\prime$. 
In the spirit of the standard quasistationary description,\cite{landauBOOKv5} we expand only up to the linear order in the nonequilibrium quantities $J$ and $\partial_t\mathbf{m}$, so that terms of the form, e.g., $J^2\partial_t\mathbf{m}\cdot\partial_x\mathbf{m}$ are excluded. To the second order in $\partial_x\mathbf{m}$, the only possible terms satisfying these requirements are:
\begin{align}
\Delta\mathcal{E}^\prime-R^\prime J&=\nonumber\\
&\hspace{-1cm}\beta\mathcal{P}\oint dx\,\partial_x\mathbf{m}\cdot\partial_t\mathbf{m}-\eta\frac{\beta^2\mathcal{P}^2}{\alpha S}J\oint dx(\partial_x\mathbf{m})^2\,.
\label{dE}
\end{align}
The first term stems physically from a spin mistracking of electrons propagating through the magnetic texture.\cite{tserkovPRB08} Since the mistracking should scale as $1/\Delta_{\rm xc}$ (vanishing in the limit of infinite exchange $\Delta_{\rm xc}$), we may anticipate the dissipative coupling to be governed by a small parameter $\beta\sim\hbar/\tau_s\Delta_{\rm xc}$, where $\tau_s$ is a characteristic (transverse) spin-dephasing time. The $\eta$ term in Eq.~(\ref{ohmD}) describes the resistance associated with magnetic texture, which is often discussed in the context of magnetic domain walls.\cite{viretPRB96} Both terms in Eq.~(\ref{dE}) are odd under time reversal, like ohmic resistance and Gilbert damping. Finally, we note that including in Eq.~(\ref{dE}) a reactive term of the form (\ref{EMF}) would not add anything new to the following considerations, as long as we treat $\mathcal{P}$ as a phenomenological coefficient.

Our modification of Ohm's law must respect the Onsager reciprocity principle.\cite{landauBOOKv5} Substituting $\partial_t\mathbf{m}$ from Eqs.~(\ref{LL}) into Eq.~(\ref{dE}), we see how the effective field $\mathbf{H}$ (which is conjugate to $\mathbf{m}$) affects the dynamics of $\mathcal{J}$. The Onsager theorem is now readily applied to determine how the electric current $J$ (which is conjugate to $\mathcal{J}$) should modify the dynamics of $\mathbf{m}$.  We write the final result as a correction to the spin torque (\ref{tau}):
\begin{equation}
\Delta\boldsymbol{\tau}^\prime=\beta\mathcal{P}J\mathbf{m}\times\partial_x\mathbf{m}\,.
\label{tauD}
\end{equation}
The complete equation of motion of the magnetic texture in the LLG form thus becomes
\begin{equation}
\partial_t\mathbf{m}=\mathbf{H}\times\mathbf{m}/S-\alpha\mathbf{m}\times\partial_t\mathbf{m}+\Delta\boldsymbol{\tau}^\prime/S\,,
\label{LLGD}
\end{equation}
with $\boldsymbol{\tau}^\prime$ implicitly included in $\mathbf{H}$.  

Eqs.~(\ref{ohmD}) and (\ref{LLGD}) are our final coupled deterministic equations. We can rewrite them in a more explicit form as
\begin{align}
L\partial_tJ+(R+R^\prime)J+\partial_t\Phi/c&=\nonumber\\
&\hspace{-1.5cm}\mathcal{P}\oint dx\,\partial_x\mathbf{m}\cdot(\beta-\mathbf{m}\times)\partial_t\mathbf{m}\,,\nonumber\\
S(1+\alpha\mathbf{m}\times)\partial_t\mathbf{m}+\mathbf{m}\times\mathbf{H}&=\mathcal{P}J(1+\beta\mathbf{m}\times)\partial_x\mathbf{m}\,.
\label{fm}
\end{align}
Here, the deterministic spin-torque contribution (\ref{tau}) is for clarity separated out of the effective field $\mathbf{H}$, which here consists of the usual purely magnetic contributions. The left-hand sides in these equations contain the ordinary Ohm's law (corrected for the magnetic-texture resistance $R^\prime$) and the LLG terms, respectively, while the right-hand sides describe the reactive Berry-phase coupling and its dissipative $\beta$ correction.

Eq.~(\ref{tauD}) was derived microscopically in Refs.~\onlinecite{zhangPRL04,tserkovPRB06md,kohnoJPSJ06,duinePRB07fk}, relating $\beta$ to electron spin dephasing: $\beta\sim\hbar/\tau_s\Delta_{\rm xc}$ (consistent with our anticipation above). Its Onsager counterpart in Eq.~(\ref{dE}) was first obtained phenomenologically in Ref.~\onlinecite{tserkovPRB08} and microscopically in Ref.~\onlinecite{duinePRB08}. These ``$\beta$ terms" are now accepted to be crucial in understanding current-driven magnetic dynamics and the reciprocal gauge fields.

\section{Dissipation power}

Suppose we perturb our system with some nonequilibrium current and magnetic texture, after which the system evolves back toward equilibrium according to the equations of motion, producing entropy. If the system is steadily driven, the heat will be dissipated to the environment at some finite rate. From standard thermodynamics, the dissipation power is  
\begin{widetext}
\begin{equation}
P[\mathbf{m}(x,t),J(t)]\equiv-J\partial_t\mathcal{J}-\oint dx\,\mathbf{H}\cdot\partial_t\mathbf{m}=RJ^2+\oint dx\left[\alpha S(\partial_t\mathbf{m})^2-2\beta\mathcal{P}J\partial_x\mathbf{m}\cdot\partial_t\mathbf{m}+\eta\frac{\beta^2\mathcal{P}^2}{\alpha S}J^2(\partial_x\mathbf{m})^2\right]\,.
\label{P}
\end{equation}
\end{widetext}
According to the second law of thermodynamics, the dissipation $(\ref{P})$ must always be positive, which means that $\eta\geq1$. This gives us the lower bound on the resistivity of the magnetic texture:
\begin{equation}
\rho=\eta\frac{\beta^2\mathcal{P}^2}{\alpha S}(\partial_x\mathbf{m})^2\geq\frac{\beta^2\mathcal{P}^2}{\alpha S}(\partial_x\mathbf{m})^2\,.
\label{rho}
\end{equation}

In models where $\alpha$ comes solely from the coupling of the magnetization to the conducting electrons (which is in fact believed to be the dominant cause for Gilbert damping in metallic ferromagnets), we may expect the lower bound (\ref{rho}) to give an estimate for the texture resistivity. For a mean-field Stoner-model treatment of Gilbert damping, we found $\alpha=\beta$, while for an $s-d$ model we had $\alpha=(s/S)\beta$, where $s$ is the portion of spin density carried by the $s$ electrons, $S$ is the total spin density, and $\beta=\hbar/\tau_s\Delta_{\rm xc}$ in both cases (with the spin-dephasing time $\tau_s$ governed by the magnetic and spin-orbit impurities).\cite{tserkovPRB06md} In both models, therefore, $\alpha S=s\beta$, giving for the resistivity estimate (up to the second order in spatial derivative)
\begin{equation}
\rho\gtrsim(\beta\mathcal{P}^2/s)(\partial_x\mathbf{m})^2\,,
\label{rha}
\end{equation}
which involves only quantities related to conducting electrons. Taking parameters relevant to Permalloy wires:\cite{yamaguchiPRL04} $p\sim1$, $\beta\sim10^{-2}$, domain-wall width of 20~nm, and the magnetization of $10^3~{\rm emu}/{\rm cm}^3$, we find the resistivity (\ref{rha}) to be $\rho\sim10^{-4}~\mu\Omega\cdot$cm. This is smaller than the domain-wall resistivity calculated to the $(1/\Delta_{\rm xc})^2$ order in spin mistracking of the magnetic profile (but still quadratic order in texture), in the absence of spin relaxation,\cite{viretPRB96} whose overall prefactor appears to be larger than in our Eq.~(\ref{rha}) for transition metals. We thus conclude that our $\eta$ may in practice be much larger than unity (which is the lower bound necessary for the consistency of our phenomenology).

Let us also note in the passing that in the special case of $\alpha=\beta$ and $\eta=1$, the magnetic dissipation (\ref{P}) acquires a very simple form:
\begin{equation}
P[\mathbf{m}(x,t)]\to\alpha S\oint dx\left(\partial_t\mathbf{m}-\frac{\mathcal{P
}J}{S}\partial_x\mathbf{m}\right)^2\,,
\end{equation}
which is nothing but the Gilbert dissipation with the advective time derivative $D_t=\partial_t+v\partial_x$ ($v=-\mathcal{P}J/S$). It is clear that this limit describes dissipative magnetic dynamics that are simply carried by the electric flow at speed $v$. In this case, the spin torques disappear if we write the LLG equation (\ref{LLG}) with $D_t$ in the place of $\partial_t$ .\cite{barnesPRL05}

\section{Thermal noise}

At finite temperatures, thermal agitation causes fluctuations of the current and magnetization, which are correlated due to their coupling. A complete description requires that we supplement the stochastic equations of motion with the correlators of these fluctuations.  It is convenient to regard these fluctuations as being due to a stochastic external magnetic field $\delta\mathbf{h}$ and a stochastic current source $\delta J$: their noise correlators are then related to the dissipative coefficients of the theory according to the fluctuation-dissipation theorem (FDT). Constructing the noise sources by following the standard procedure,\cite{landauBOOKv5} our final coupled stochastic equations become:
\begin{widetext}
\begin{align}
\label{fe}
L\partial_tJ+\tilde{R}(J+\delta J)+\partial_t\Phi/c&=\mathcal{P}\oint dx\,\partial_x\mathbf{m}\cdot(\beta-\mathbf{m}\times)\partial_t\mathbf{m}\,,\\
S(1+\alpha\mathbf{m}\times)\partial_t\mathbf{m}+\mathbf{m}\times(\mathbf{H}+\delta\mathbf{h})&=\mathcal{P}J\partial_x\mathbf{m}+\mathcal{P}(J+\delta J)\beta\mathbf{m}\times\partial_x\mathbf{m}\,,
\label{fm}
\end{align}
where we have explicitly separated the deterministic spin-torque contribution $\mathcal{P}J\partial_x\mathbf{m}$ out of the effective field $\mathbf{H}$, which here consists of the usual purely magnetic contributions. The left-hand sides in these equations contain the ordinary Ohm's law (corrected for the magnetic-texture resistance: $\tilde{R}=R+R^\prime$) and the LLG terms, respectively, while the right-hand sides describe the reactive Berry-phase coupling and its dissipative $\beta$ correction. 

Writing $\{J,\mathbf{H}\}=-\hat{\gamma}\otimes\{\partial_t\mathcal{J},\partial_t\mathbf{m}\}$, we read out the ``matrix" $\hat{\gamma}$ from Eqs.~(\ref{fe}) and (\ref{fm}):\begin{align}
\hat{\gamma}_{J,J}&=\frac{1}{R^\prime}\,,\,\,\,\,\,\,\hat{\gamma}_{J,\mathbf{h}(x)}=-\frac{\beta\mathcal{P}}{R^\prime}\partial_x\mathbf{m}\,,\,\,\,\,\,\,\hat{\gamma}_{\mathbf{h}(x),J}=\frac{\beta\mathcal{P}}{R^\prime}\partial_x\mathbf{m}\,,\nonumber\\
\hat{\gamma}_{h_i(x),h_{i^\prime}(x^\prime)}&=S\epsilon^{ii^\prime j}m_j(x)\delta(x-x^\prime)+\alpha S\delta_{ii^\prime}\delta(x-x^\prime)-\frac{\beta^2\mathcal{P}^2}{R^\prime}\partial_xm_i(x)\partial_xm_{i^\prime}(x^\prime)
\end{align}
where $\epsilon^{ijk}$ is the antisymmetric Levi-Civita tensor. Symmetrizing matrix $\hat{\gamma}$ immediately produces Langevin sources satisfying the FDT,\cite{landauBOOKv5} in the limit that $ \hbar \omega \ll k_B T$:
\begin{align}
\langle\delta J(t)\delta J(t^\prime)\rangle&=2k_BT\delta(t-t^\prime)/\tilde{R}\,,\,\,\,\langle\delta J(t)\delta\mathbf{h}(t^\prime)\rangle=0\,,\nonumber\\
\langle\delta h_i(x)\delta h_{i^\prime}(x^\prime)\rangle&=2k_BT\left[\alpha S\delta_{ii^\prime}\delta(x-x^\prime)-(\beta^2\mathcal{P}^2/\tilde{R})\partial_xm_i\partial_{x^\prime}m_{i^\prime}\right]\delta(t-t^\prime)\,.
\end{align}
Apart from the obvious contributions, we have a magnetic field noise proportional to $\beta^2$, in the form of a nonlocal tensor Gilbert damping. The nonlocal Gilbert damping is apparent, if the electrons are not externally driven, $\partial_t\Phi=0$, in the limit $L\to0$ of a large ring, in which case the magnetic equation decouples to give
\begin{equation}
S(1+\alpha\mathbf{m}\times)\partial_t\mathbf{m}+\mathbf{m}\times(\mathbf{H}+\delta\mathbf{h}+\delta\mathbf{h}^\prime)=\frac{\mathcal{P}^2}{\tilde{R}}(1+\beta\mathbf{m}\times)\partial_x\mathbf{m}\oint dx^\prime\,\partial_{x^\prime}\mathbf{m}\cdot(\beta-\mathbf{m}\times)\partial_t\mathbf{m}\,.
\label{mmm}
\end{equation}
Here, we moved the spin torque driven by the Nyquist noise to the left as
\begin{equation}
\delta\mathbf{h}^\prime=-\mathcal{P}\delta J\mathbf{m}\times\partial_x\mathbf{m}\,.
\end{equation}
$\delta\mathbf{h}^\prime$ thus enters the equation as a statistically independent current-driven noise source. Writing the right-hand side of Eq.~(\ref{mmm}) as
\begin{equation}
-\mathbf{m}\times\oint dx^\prime\tensor{\mathbf{K}}(x,x^\prime)\partial_t\mathbf{m}(x^\prime)\,,
\end{equation}
where
\begin{equation}
K_{ii^\prime}(x,x^\prime)=\frac{\mathcal{P}^2}{R^\prime}\left(\mathbf{m}\times\partial_x\mathbf{m}-\beta\partial_x\mathbf{m}\right)_i\left(\mathbf{m}\times\partial_{x^\prime}\mathbf{m}+\beta\partial_{x^\prime}\mathbf{m}\right)_{i^\prime}\,,
\end{equation}
and extracting the symmetric part of the tensor $K_{ii^\prime}(x,x^\prime)$, we arrive at the total Gilbert damping tensor
\begin{equation}
G_{ii^\prime}(x,x^\prime)=\alpha\delta_{ii^\prime}\delta(x-x^\prime)+\frac{\mathcal{P}^2}{S\tilde{R}}\left[(\mathbf{m}\times\partial_x\mathbf{m})_i(\mathbf{m}\times\partial_{x^\prime}\mathbf{m})_{i^\prime}-\beta^2\partial_xm_i\partial_{x^\prime}m_{i^\prime}\right]\,.
\label{G}
\end{equation}
\end{widetext}
This is exactly the form required by the FDT, consistent with the correlator for $\delta\mathbf{h}+\delta\mathbf{h}^\prime$. The effective Gilbert damping can thus appear both negative and positive in different regions. The minimal texture resistivity (\ref{rho}), however, insures that we have a nonnegative damping globally. This Gilbert damping originates physically in the spin torques that are generated by the magnetically-driven fictitious EMF. Nonlocal $\partial_x\partial_{x^\prime}$ magnetic noise was recently constructed in Ref.~\onlinecite{forosPRB08} (neglecting spin relaxation and $\beta$) by heuristically converting Nyquist current noise into magnetic fluctuations via adiabatic spin transfer. Although the DFT-required nonlocal $\partial_x\partial_{x^\prime}$ Gilbert tensor (\ref{G}) was established in that paper (apart from the $\beta^2$ piece), only here we are able to derive it directly from the fundamental Langevin sources of the coupled magnetohydrodynamic theory, dictated by the FDT. As estimated in Ref.~\onlinecite{forosPRB08}, this nonlocal contribution to Gilbert damping is in practice important (in comparison to $\alpha$) in nanoscale magnetic structures.

\section{Summary}

We developed a general phenomenological theory of magnetohydrodynamic coupling in isotropic metallic ferromagnets. The reactive coupling between magnetic texture dynamics on the one hand and electric flows on the other stems from the Berry phase accumulated by electron spin following the quasistationary magnetic texture. Dissipative terms of the coupled dynamic equations originate in the electron spin mistracking of the magnetic order parameter and the associated spin dephasing. Apart from the usual Gilbert damping, the latter leads to a viscous coupling between electric currents and magnetic texture dynamics, parametrized by a single parameter $\beta$. We also obtain a small correction to the texture resistivity at order $\beta^2$. Finally, our thermodynamic description of the magnetohydrodynamic coupling allows us to derive the stochastic Langevin contributions to the effective field and electric current, according to the fluctuation-dissipation theorem.

\acknowledgements

We acknowledge stimulating discussions with Gerrit E. W. Bauer, Arne Brataas, and Mark D. Stiles. This work was supported in part by the Alfred P. Sloan Foundation.


\begin{thebibliography}{35}
\expandafter\ifx\csname natexlab\endcsname\relax\def\natexlab#1{#1}\fi
\expandafter\ifx\csname bibnamefont\endcsname\relax
  \def\bibnamefont#1{#1}\fi
\expandafter\ifx\csname bibfnamefont\endcsname\relax
  \def\bibfnamefont#1{#1}\fi
\expandafter\ifx\csname citenamefont\endcsname\relax
  \def\citenamefont#1{#1}\fi
\expandafter\ifx\csname url\endcsname\relax
  \def\url#1{\texttt{#1}}\fi
\expandafter\ifx\csname urlprefix\endcsname\relax\def\urlprefix{URL }\fi
\providecommand{\bibinfo}[2]{#2}
\providecommand{\eprint}[2][]{\url{#2}}

\bibitem[{\citenamefont{Berry}(1984)}]{berryPRSLA84}
\bibinfo{author}{\bibfnamefont{M.~V.} \bibnamefont{Berry}},
  \bibinfo{journal}{Proc. R. Soc. London A} \textbf{\bibinfo{volume}{392}},
  \bibinfo{pages}{45} (\bibinfo{year}{1984}).

\bibitem[{\citenamefont{Volovik}(1987)}]{volovikJPC87}
\bibinfo{author}{\bibfnamefont{G.~E.} \bibnamefont{Volovik}},
  \bibinfo{journal}{J. Phys. C: Sol. State Phys.}
  \textbf{\bibinfo{volume}{20}}, \bibinfo{pages}{L83} (\bibinfo{year}{1987}).

\bibitem[{\citenamefont{Landau et~al.}(1984)\citenamefont{Landau, Lifshitz, and
  Pitaevskii}}]{landauBOOKv8}
\bibinfo{author}{\bibfnamefont{L.~D.} \bibnamefont{Landau}},
  \bibinfo{author}{\bibfnamefont{E.~M.} \bibnamefont{Lifshitz}},
  \bibnamefont{and} \bibinfo{author}{\bibfnamefont{L.~P.}
  \bibnamefont{Pitaevskii}}, \emph{\bibinfo{title}{Electrodynamics of
  Continuous Media}}, vol.~\bibinfo{volume}{8} of \emph{\bibinfo{series}{Course
  of Theoretical Physics}} (\bibinfo{publisher}{Pergamon},
  \bibinfo{address}{Oxford}, \bibinfo{year}{1984}), \bibinfo{edition}{2nd} ed.

\bibitem[{\citenamefont{Zhang and Li}(2004)}]{zhangPRL04}
\bibinfo{author}{\bibfnamefont{S.}~\bibnamefont{Zhang}} \bibnamefont{and}
  \bibinfo{author}{\bibfnamefont{Z.}~\bibnamefont{Li}}, \bibinfo{journal}{Phys.
  Rev. Lett.} \textbf{\bibinfo{volume}{93}}, \bibinfo{eid}{127204}
  (\bibinfo{year}{2004}).

\bibitem[{\citenamefont{Thiaville et~al.}(2005)\citenamefont{Thiaville,
  Nakatani, Miltat, and Suzuki}}]{thiavilleEPL05}
\bibinfo{author}{\bibfnamefont{A.}~\bibnamefont{Thiaville}},
  \bibinfo{author}{\bibfnamefont{Y.}~\bibnamefont{Nakatani}},
  \bibinfo{author}{\bibfnamefont{J.}~\bibnamefont{Miltat}}, \bibnamefont{and}
  \bibinfo{author}{\bibfnamefont{Y.}~\bibnamefont{Suzuki}},
  \bibinfo{journal}{Europhys. Lett.} \textbf{\bibinfo{volume}{69}},
  \bibinfo{pages}{990} (\bibinfo{year}{2005}).

\bibitem[{\citenamefont{Tserkovnyak et~al.}(2006)\citenamefont{Tserkovnyak,
  Skadsem, Brataas, and Bauer}}]{tserkovPRB06md}
\bibinfo{author}{\bibfnamefont{Y.}~\bibnamefont{Tserkovnyak}},
  \bibinfo{author}{\bibfnamefont{H.~J.} \bibnamefont{Skadsem}},
  \bibinfo{author}{\bibfnamefont{A.}~\bibnamefont{Brataas}}, \bibnamefont{and}
  \bibinfo{author}{\bibfnamefont{G.~E.~W.} \bibnamefont{Bauer}},
  \bibinfo{journal}{Phys. Rev. B} \textbf{\bibinfo{volume}{74}},
  \bibinfo{eid}{144405} (\bibinfo{year}{2006});
  \bibinfo{author}{\bibfnamefont{H.~J.} \bibnamefont{Skadsem}},
  \bibinfo{author}{\bibfnamefont{Y.}~\bibnamefont{Tserkovnyak}},
  \bibinfo{author}{\bibfnamefont{A.}~\bibnamefont{Brataas}}, \bibnamefont{and}
  \bibinfo{author}{\bibfnamefont{G.~E.~W.} \bibnamefont{Bauer}},
  \textit{ibid.} \textbf{\bibinfo{volume}{75}},
  \bibinfo{eid}{094416} (\bibinfo{year}{2007});
  \bibinfo{author}{\bibfnamefont{Y.}~\bibnamefont{Tserkovnyak}},
  \bibinfo{author}{\bibfnamefont{A.}~\bibnamefont{Brataas}}, \bibnamefont{and}
  \bibinfo{author}{\bibfnamefont{G.~E.} \bibnamefont{Bauer}},
  \bibinfo{journal}{J. Magn. Magn. Mater.} \textbf{\bibinfo{volume}{320}},
  \bibinfo{pages}{1282} (\bibinfo{year}{2008}).

\bibitem[{\citenamefont{Yamaguchi et~al.}(2004)\citenamefont{Yamaguchi, Ono,
  Nasu, Miyake, Mibu, and Shinjo}}]{yamaguchiPRL04}
\bibinfo{author}{\bibfnamefont{A.}~\bibnamefont{Yamaguchi}},
  \bibinfo{author}{\bibfnamefont{T.}~\bibnamefont{Ono}},
  \bibinfo{author}{\bibfnamefont{S.}~\bibnamefont{Nasu}},
  \bibinfo{author}{\bibfnamefont{K.}~\bibnamefont{Miyake}},
  \bibinfo{author}{\bibfnamefont{K.}~\bibnamefont{Mibu}}, \bibnamefont{and}
  \bibinfo{author}{\bibfnamefont{T.}~\bibnamefont{Shinjo}},
  \bibinfo{journal}{Phys. Rev. Lett.} \textbf{\bibinfo{volume}{92}},
  \bibinfo{eid}{077205} (\bibinfo{year}{2004});
  \bibinfo{author}{\bibfnamefont{M.}~\bibnamefont{Hayashi}},
  \bibinfo{author}{\bibfnamefont{L.}~\bibnamefont{Thomas}},
  \bibinfo{author}{\bibfnamefont{Y.~B.} \bibnamefont{Bazaliy}},
  \bibinfo{author}{\bibfnamefont{C.}~\bibnamefont{Rettner}},
  \bibinfo{author}{\bibfnamefont{R.}~\bibnamefont{Moriya}},
  \bibinfo{author}{\bibfnamefont{X.}~\bibnamefont{Jiang}}, \bibnamefont{and}
  \bibinfo{author}{\bibfnamefont{S.~S.~P.} \bibnamefont{Parkin}},
  \textit{ibid.} \textbf{\bibinfo{volume}{96}},
  \bibinfo{eid}{197207} (\bibinfo{year}{2006});
  \bibinfo{author}{\bibfnamefont{M.}~\bibnamefont{Hayashi}},
  \bibinfo{author}{\bibfnamefont{L.}~\bibnamefont{Thomas}},
  \bibinfo{author}{\bibfnamefont{C.}~\bibnamefont{Rettner}},
  \bibinfo{author}{\bibfnamefont{R.}~\bibnamefont{Moriya}}, \bibnamefont{and}
  \bibinfo{author}{\bibfnamefont{S.~S.~P.} \bibnamefont{Parkin}},
  \bibinfo{journal}{Nature Phys.} \textbf{\bibinfo{volume}{3}},
  \bibinfo{pages}{21} (\bibinfo{year}{2007}).

\bibitem[{\citenamefont{Barnes and Maekawa}(2005)}]{barnesPRL05}
\bibinfo{author}{\bibfnamefont{S.~E.} \bibnamefont{Barnes}} \bibnamefont{and}
  \bibinfo{author}{\bibfnamefont{S.}~\bibnamefont{Maekawa}},
  \bibinfo{journal}{Phys. Rev. Lett.} \textbf{\bibinfo{volume}{95}},
  \bibinfo{eid}{107204} (\bibinfo{year}{2005}).

\bibitem[{\citenamefont{Yamanouchi et~al.}(2007)\citenamefont{Yamanouchi, Ieda,
  Matsukura, Barnes, Maekawa, and Ohno}}]{yamanouchiSCI07}
\bibinfo{author}{\bibfnamefont{M.}~\bibnamefont{Yamanouchi}},
  \bibinfo{author}{\bibfnamefont{J.}~\bibnamefont{Ieda}},
  \bibinfo{author}{\bibfnamefont{F.}~\bibnamefont{Matsukura}},
  \bibinfo{author}{\bibfnamefont{S.~E.} \bibnamefont{Barnes}},
  \bibinfo{author}{\bibfnamefont{S.}~\bibnamefont{Maekawa}}, \bibnamefont{and}
  \bibinfo{author}{\bibfnamefont{H.}~\bibnamefont{Ohno}},
  \bibinfo{journal}{Science} \textbf{\bibinfo{volume}{317}},
  \bibinfo{pages}{1726} (\bibinfo{year}{2007}).

\bibitem[{\citenamefont{Kohno and Shibata}(2007)}]{kohnoJPSJ07}
\bibinfo{author}{\bibfnamefont{H.}~\bibnamefont{Kohno}} \bibnamefont{and}
  \bibinfo{author}{\bibfnamefont{J.}~\bibnamefont{Shibata}},
  \bibinfo{journal}{J. Phys. Soc. Jpn.} \textbf{\bibinfo{volume}{76}},
  \bibinfo{pages}{063710} (\bibinfo{year}{2007}).

\bibitem[{\citenamefont{Barnes and Maekawa}(2007)}]{barnesPRL07}
\bibinfo{author}{\bibfnamefont{S.~E.} \bibnamefont{Barnes}} \bibnamefont{and}
  \bibinfo{author}{\bibfnamefont{S.}~\bibnamefont{Maekawa}},
  \bibinfo{journal}{Phys. Rev. Lett.} \textbf{\bibinfo{volume}{98}},
  \bibinfo{eid}{246601} (\bibinfo{year}{2007}).

\bibitem[{\citenamefont{Saslow}(2007)}]{saslowPRB07}
\bibinfo{author}{\bibfnamefont{W.~M.} \bibnamefont{Saslow}},
  \bibinfo{journal}{Phys. Rev. B} \textbf{\bibinfo{volume}{76}},
  \bibinfo{eid}{184434} (\bibinfo{year}{2007});
  \bibinfo{author}{\bibfnamefont{S.~A.} \bibnamefont{Yang}},
  \bibinfo{author}{\bibfnamefont{D.}~\bibnamefont{Xiao}}, \bibnamefont{and}
  \bibinfo{author}{\bibfnamefont{Q.}~\bibnamefont{Niu}},
  \bibinfo{note}{cond-mat/0709.1117}.

\bibitem[{\citenamefont{Duine}(2008)}]{duinePRB08}
\bibinfo{author}{\bibfnamefont{R.~A.} \bibnamefont{Duine}},
  \bibinfo{journal}{Phys. Rev. B} \textbf{\bibinfo{volume}{77}},
  \bibinfo{eid}{014409} (\bibinfo{year}{2008}).

\bibitem[{\citenamefont{Tserkovnyak and Mecklenburg}(2008)}]{tserkovPRB08}
\bibinfo{author}{\bibfnamefont{Y.}~\bibnamefont{Tserkovnyak}} \bibnamefont{and}
  \bibinfo{author}{\bibfnamefont{M.}~\bibnamefont{Mecklenburg}},
  \bibinfo{journal}{Phys. Rev. B} \textbf{\bibinfo{volume}{77}},
  \bibinfo{eid}{134407} (\bibinfo{year}{2008}).

\bibitem[{\citenamefont{Moriyama et~al.}(2008)\citenamefont{Moriyama, Cao, Fan,
  Xuan, Nikoli{\'c}, Tserkovnyak, Kolodzey, and Xiao}}]{moriyamaPRL08}
\bibinfo{author}{\bibfnamefont{T.}~\bibnamefont{Moriyama}},
  \bibinfo{author}{\bibfnamefont{R.}~\bibnamefont{Cao}},
  \bibinfo{author}{\bibfnamefont{X.}~\bibnamefont{Fan}},
  \bibinfo{author}{\bibfnamefont{G.}~\bibnamefont{Xuan}},
  \bibinfo{author}{\bibfnamefont{B.~K.} \bibnamefont{Nikoli{\'c}}},
  \bibinfo{author}{\bibfnamefont{Y.}~\bibnamefont{Tserkovnyak}},
  \bibinfo{author}{\bibfnamefont{J.}~\bibnamefont{Kolodzey}}, \bibnamefont{and}
  \bibinfo{author}{\bibfnamefont{J.~Q.} \bibnamefont{Xiao}},
  \bibinfo{journal}{Phys. Rev. Lett.} \textbf{\bibinfo{volume}{100}},
  \bibinfo{eid}{067602} (\bibinfo{year}{2008}).

\bibitem[{\citenamefont{Tserkovnyak et~al.}()\citenamefont{Tserkovnyak,
  Moriyama, and Xiao}}]{tserkovCM08tb}
\bibinfo{author}{\bibfnamefont{Y.}~\bibnamefont{Tserkovnyak}},
  \bibinfo{author}{\bibfnamefont{T.}~\bibnamefont{Moriyama}}, \bibnamefont{and}
  \bibinfo{author}{\bibfnamefont{J.~Q.} \bibnamefont{Xiao}},
  \bibinfo{journal}{Phys. Rev. B} \textbf{\bibinfo{volume}{78}},
  \bibinfo{eid}{020401(R)} (\bibinfo{year}{2008}).

\bibitem[{\citenamefont{Tserkovnyak et~al.}(2005)\citenamefont{Tserkovnyak,
  Brataas, Bauer, and Halperin}}]{tserkovRMP05}
\bibinfo{author}{\bibfnamefont{Y.}~\bibnamefont{Tserkovnyak}},
  \bibinfo{author}{\bibfnamefont{A.}~\bibnamefont{Brataas}},
  \bibinfo{author}{\bibfnamefont{G.~E.~W.} \bibnamefont{Bauer}},
  \bibnamefont{and} \bibinfo{author}{\bibfnamefont{B.~I.}
  \bibnamefont{Halperin}}, \bibinfo{journal}{Rev. Mod. Phys.}
  \textbf{\bibinfo{volume}{77}}, \bibinfo{eid}{1375} (\bibinfo{year}{2005});
  \bibinfo{author}{\bibfnamefont{A.}~\bibnamefont{Brataas}},
  \bibinfo{author}{\bibfnamefont{G.~E.~W.} \bibnamefont{Bauer}},
  \bibnamefont{and} \bibinfo{author}{\bibfnamefont{P.~J.} \bibnamefont{Kelly}},
  \bibinfo{journal}{Phys. Rep.} \textbf{\bibinfo{volume}{427}},
  \bibinfo{pages}{157} (\bibinfo{year}{2006});
  \bibinfo{author}{\bibfnamefont{D.~C.} \bibnamefont{Ralph}} \bibnamefont{and}
  \bibinfo{author}{\bibfnamefont{M.~D.} \bibnamefont{Stiles}},
  \bibinfo{journal}{J. Magn. Magn. Mater.} \textbf{\bibinfo{volume}{320}},
  \bibinfo{pages}{1190} (\bibinfo{year}{2007}).

\bibitem[{\citenamefont{Stern}(1992)}]{sternPRL92}
\bibinfo{author}{\bibfnamefont{A.}~\bibnamefont{Stern}},
  \bibinfo{journal}{Phys. Rev. Lett.} \textbf{\bibinfo{volume}{68}},
  \bibinfo{pages}{1022} (\bibinfo{year}{1992});
  \bibinfo{author}{\bibfnamefont{Y.}~\bibnamefont{Aharonov}} \bibnamefont{and}
  \bibinfo{author}{\bibfnamefont{A.}~\bibnamefont{Stern}},
  \bibinfo{journal}{Phys. Rev. Lett.} \textbf{\bibinfo{volume}{69}},
  \bibinfo{pages}{3593} (\bibinfo{year}{1992}).

\bibitem[{\citenamefont{Lifshitz and Pitaevskii}(1980)}]{landauBOOKv9}
\bibinfo{author}{\bibfnamefont{E.~M.} \bibnamefont{Lifshitz}} \bibnamefont{and}
  \bibinfo{author}{\bibfnamefont{L.~P.} \bibnamefont{Pitaevskii}},
  \emph{\bibinfo{title}{Statistical Physics, Part 2}}, vol.~\bibinfo{volume}{9}
  of \emph{\bibinfo{series}{Course of Theoretical Physics}}
  (\bibinfo{publisher}{Pergamon}, \bibinfo{address}{Oxford},
  \bibinfo{year}{1980}), \bibinfo{edition}{3rd} ed.

\bibitem[{\citenamefont{Gilbert}(2004)}]{gilbertIEEEM04}
\bibinfo{author}{\bibfnamefont{T.~L.} \bibnamefont{Gilbert}},
  \bibinfo{journal}{IEEE Trans. Magn.} \textbf{\bibinfo{volume}{40}},
  \bibinfo{pages}{3443} (\bibinfo{year}{2004}).

\bibitem[{\citenamefont{Bazaliy et~al.}(1998)\citenamefont{Bazaliy, Jones, and
  Zhang}}]{bazaliyPRB98}
\bibinfo{author}{\bibfnamefont{Y.~B.} \bibnamefont{Bazaliy}},
  \bibinfo{author}{\bibfnamefont{B.~A.} \bibnamefont{Jones}}, \bibnamefont{and}
  \bibinfo{author}{\bibfnamefont{S.-C.} \bibnamefont{Zhang}},
  \bibinfo{journal}{Phys. Rev. B} \textbf{\bibinfo{volume}{57}},
  \bibinfo{pages}{R3213} (\bibinfo{year}{1998}).

\bibitem[{\citenamefont{Landau and Lifshitz}(1980)}]{landauBOOKv5}
\bibinfo{author}{\bibfnamefont{L.~D.} \bibnamefont{Landau}} \bibnamefont{and}
  \bibinfo{author}{\bibfnamefont{E.~M.} \bibnamefont{Lifshitz}},
  \emph{\bibinfo{title}{Statistical Physics, Part 1}}, vol.~\bibinfo{volume}{5}
  of \emph{\bibinfo{series}{Course of Theoretical Physics}}
  (\bibinfo{publisher}{Pergamon}, \bibinfo{address}{Oxford},
  \bibinfo{year}{1980}), \bibinfo{edition}{3rd} ed.

\bibitem[{\citenamefont{Viret et~al.}(1996)\citenamefont{Viret, Vignoles, Cole,
  Coey, Allen, Daniel, and Gregg}}]{viretPRB96}
\bibinfo{author}{\bibfnamefont{M.}~\bibnamefont{Viret}},
  \bibinfo{author}{\bibfnamefont{D.}~\bibnamefont{Vignoles}},
  \bibinfo{author}{\bibfnamefont{D.}~\bibnamefont{Cole}},
  \bibinfo{author}{\bibfnamefont{J.~M.~D.} \bibnamefont{Coey}},
  \bibinfo{author}{\bibfnamefont{W.}~\bibnamefont{Allen}},
  \bibinfo{author}{\bibfnamefont{D.~S.} \bibnamefont{Daniel}},
  \bibnamefont{and} \bibinfo{author}{\bibfnamefont{J.~F.} \bibnamefont{Gregg}},
  \bibinfo{journal}{Phys. Rev. B} \textbf{\bibinfo{volume}{53}},
  \bibinfo{pages}{8464} (\bibinfo{year}{1996});
  \bibinfo{author}{\bibfnamefont{C.~H.} \bibnamefont{Marrows}},
  \bibinfo{journal}{Adv. Phys.} \textbf{\bibinfo{volume}{54}},
  \bibinfo{pages}{585} (\bibinfo{year}{2005}).

\bibitem[{\citenamefont{Kohno et~al.}(2006)\citenamefont{Kohno, Tatara, and
  Shibata}}]{kohnoJPSJ06}
\bibinfo{author}{\bibfnamefont{H.}~\bibnamefont{Kohno}},
  \bibinfo{author}{\bibfnamefont{G.}~\bibnamefont{Tatara}}, \bibnamefont{and}
  \bibinfo{author}{\bibfnamefont{J.}~\bibnamefont{Shibata}},
  \bibinfo{journal}{J. Phys. Soc. Jpn.} \textbf{\bibinfo{volume}{75}},
  \bibinfo{pages}{113706} (\bibinfo{year}{2006}).

\bibitem[{\citenamefont{Duine et~al.}(2007)\citenamefont{Duine, N{\'u}{\~n}ez,
  Sinova, and MacDonald}}]{duinePRB07fk}
\bibinfo{author}{\bibfnamefont{R.~A.} \bibnamefont{Duine}},
  \bibinfo{author}{\bibfnamefont{A.~S.} \bibnamefont{N{\'u}{\~n}ez}},
  \bibinfo{author}{\bibfnamefont{J.}~\bibnamefont{Sinova}}, \bibnamefont{and}
  \bibinfo{author}{\bibfnamefont{A.~H.} \bibnamefont{MacDonald}},
  \bibinfo{journal}{Phys. Rev. B} \textbf{\bibinfo{volume}{75}},
  \bibinfo{eid}{214420} (\bibinfo{year}{2007}).

\bibitem[{\citenamefont{Foros et~al.}()\citenamefont{Foros, Brataas,
  Tserkovnyak, and Bauer}}]{forosPRB08}
\bibinfo{author}{\bibfnamefont{J.}~\bibnamefont{Foros}},
  \bibinfo{author}{\bibfnamefont{A.}~\bibnamefont{Brataas}},
  \bibinfo{author}{\bibfnamefont{Y.}~\bibnamefont{Tserkovnyak}},
  \bibnamefont{and} \bibinfo{author}{\bibfnamefont{G.~E.~W.} \bibnamefont{Bauer}},
  \bibinfo{journal}{Phys. Rev. B} \textbf{\bibinfo{volume}{78}},
  \bibinfo{pages}{140402(R)} (\bibinfo{year}{2008});
\end{thebibliography}
\end{document}